\documentclass[twocolumn,showpacs,preprintnumbers,amsmath,amssymb]{revtex4}
\usepackage{graphicx}
\usepackage{dcolumn}
\usepackage{bm}

\begin{document}

\preprint{APS/123-QED}

\title{The Phase Transition of Dark Energy}

\author{Wei Wang}
\author{Yuanxing Gui\footnote{Corresponding author, thphys@dlut.edu.cn}}%
\affiliation{%
School of Physics and Optoelectronic Technology, Dalian University
of Technology, Dalian 116024, P. R. China
}%

\author{Ying Shao}
\affiliation{ School of Physics and Optoelectronic Technology,
Dalian University of Technology, Dalian 116024, P. R. China
}%

\date{\today}

\begin{abstract}
Considering that the universe is filled with the nonrelativistic
matter and dark energy and each component is respectively
satisfied with its conservation condition in the absence of their
interaction, we give the change rate of the fractional density and
the density of dark energy from the conservation condition. It is
clear that the fractional density of dark energy will monotonously
increase and gradually become the dominating contribution to the
universe as the redshift becomes low. Combining the evolutional
trend of the state equation of dark energy and the change rate of
the density of dark energy we find that the density of dark energy
will decrease up to a minimum and whereafter it will increase
again as the redshift becomes low. This can be regarded as the
phase transition of dark energy from the quintessence phase to the
phantom phase.
\end{abstract}

 \maketitle

The recent cosmological observations strongly support that the
expansion of our universe accelerates, obtained by studying type
Ia Supernovae (SNe Ia)\cite{1}, Cosmic microwave background
(CMB)\cite{2} and large scale structure\cite{3}. This calls for a
theoretical explanation. Now the models of an exotic component of
unknown nature called ``dark energy'' with the negative pressure
are predominant in the literatures. It accounts for $\frac{2}{3}$
of the total energy of the universe and has the state equation
$\omega _{de}=\frac{p_{de}}{\rho _{de}}<-\frac{1}{3}$ to generate
acceleration in Einstein gravity. Proposals of the models of dark
energy range from the cosmological constant to the dynamical
scalar field quintessence\cite{4} (or phantom\cite{5},
quintom\cite{6}, k-essence\cite{7} and so on) and the
unified\cite{8,9} or coupled\cite{10} dark energy models have been
put forward. These models show various degrees of consistency with
the observational data in describing the evolution of the
universe. Given the plethora of models, choosing the models of the
best fit to the observational data and investigating the
properties of this mysterious component is one of the most
exciting problems of modern cosmology. The variation of the
density of dark energy with the redshift can provide a critical
clue to the nature of dark energy which have been realized in
Ref.\cite{11}. But few investigation about the evolutional trend
of the density itself of dark energy is presented in the
literatures. What is discussed in the most literatures involving
the density of dark energy is the fractional density of dark
energy.

In this paper, We perform a model-independent analysis of the
evolutional trend of the fractional density and the density of
dark energy. Expect that it can serve some guides to put forward
the fittest model of dark energy and to explore the mysterious
nature of dark energy. The analysis is based on the two following
preconditions: (1) the universe is filled with the nonrelativistic
matter and dark energy, (2) the two components have not
interaction and they are respectively satisfied with their own
conservation condition . We use the conservation condition to
derive the change rate of the density and the fractional density
of dark energy. The fractional density of dark energy increases as
the redshift becomes low. Considering the value of the state
equation of dark energy we can conclude the evolutional trend of
the density of dark energy. If choosing the preferable solution
that the state equation of dark energy evolves from $\omega
_{de}>-1$ phase to $\omega _{de}<-1$ phase,  the density of dark
energy will decrease to a minimum before increasing as the
redshift becomes low which can be regarded as the phase transition
of dark energy from the quintessence phase to the phantom phase.
The transformation point corresponds to the state equation of dark
energy $\omega _{de}=-1$. Let us see in more detail how things
work out.

We consider a spatially flat, homogeneous and isotropic universe,
is filled with two noninteracting components only, i.e., the
nonrelativistic matter and dark energy. The cosmological equations
are then
\begin{equation}
H^{2}=(\frac{\dot{a}}{a})^{2}=\frac{1}{3}\rho
_{tot}=\frac{1}{3}(\rho _{m}+\rho _{de})\,, \label{equ1}
\end{equation}
\begin{equation}
\dot{\rho}_{m}=-3H\rho _{m}\,,  \label{equ2}
\end{equation}
\begin{equation}
\dot{\rho}_{de}=-3H(\rho _{de}+p _{de})\,,  \label{equ3}
\end{equation}
where the subscripts ``m'', ``de'', ``tot'' and dot separately
denote the matter, dark energy, the total amount and the
derivative with the cosmic time. In Eq. (1) $8\pi G=1$ and the
pressure of the matter $p_{m}=0$ have been taken into account. We
can obtain the change rate of the density and the fractional
density of dark energy using the three equations above. First we
discuss the behavior of the fractional density of dark energy
which can be written as:
\begin{equation}
\Omega_{de}=\frac{\rho _{de}}{\rho _{tot}}=\frac{\rho _{de}}{(\rho
_{m}+\rho _{de})}\,. \label{equ4}
\end{equation}
Making use of the conservation equation (2), (3), we can express
the change rate of $\Omega_{de}$ as the cosmic time, as follows:
\begin{equation}
\dot{\Omega}_{de}=\frac{-3Hp _{de}\rho_{m}}{(\rho _{m}+\rho
_{de})^{2}}\,. \label{equ5}
\end{equation}
Using $H=\frac{\dot{a}}{a}$ and $\frac{a_{0}}{a}=1+z$ ($a_{0}$ is
the present scale factor, and today's evaluated quantities will
hereafter denoted by the
label $%
``0\textquotedblright $), the Eq. (5) can be changed into the
derivative with respect to the redshift $z$ (this transformation
is more convenient for our analysis):
\begin{equation}
\Omega_{de}'=\frac{d\Omega_{de}}{dz}=\frac{3p
_{de}\rho_{m}}{(1+z)(\rho _{m}+\rho _{de})^{2}}\,. \label{equ6}
\end{equation}
The variables in the Eq. (6) are satisfied with: $z>-1$, $p
_{de}<0$, $\rho _{m}>0$. So it is certain to obtain
$\Omega_{de}'<0$, which can draw a conclusion that the fractional
density of dark energy will monotonously increase as the redshift
becomes low. The result is compatible with what is described in
the many literatures. It has also a important implication for the
theory of dark energy that the model of dark energy like the
oscillating dark energy as a candidate for dark energy with no
interaction will be excluded and dark energy is sure to dominate
gradually the universe. The reverse is also true that we have to
consider the presence of interaction or the instance of the
positive pressure of dark energy to obtain the model of dark
energy of the oscillating dark energy\cite{12}.

It follows that we discuss the change rate of the density of dark
energy. We change the derivative with respect to time into the
derivative with respect to the redshift all the same. The equation
(3) can be rewritten as:
\begin{equation}
\frac{d\rho_{de}}{dz}=\rho_{de}'=\frac{3(\rho _{de}+p
_{de})}{1+z}=\frac{3\rho _{de}(1+\omega_{de})}{1+z}\,.
\label{equ7}
\end{equation}
From the equation (7) we know the sign of $1+\omega_{de}$ decides
wether the density of dark energy will increase or not as the
redshift becomes low. According to the sign of $1+\omega_{de}$, we
will divide the problems into the three instances to discuss the
variety of the density of dark energy as the redshift becomes low.

\begin{widetext}
\[
\left\{
\begin{array}{ll}
decrease  &\mbox{\quad when $1+\omega_{de}>0$ \quad like the quintessence;}\\
invariable &\mbox{\quad when $1+\omega_{de}=0$ \quad like the cosmological constant;}\\
increase   &\mbox{\quad when $1+\omega_{de}<0$ \quad  like the
phantom.}
\end{array}
\right.
\]
\end{widetext}
The present data favor an evolving dark energy with the state
equation being below $-1$ around present epoch evolved from
$\omega_{de}>-1$ in the past\cite{13}. If this is true, combining
the analysis above of the state equation of dark energy we can
draw the conclusion that in the absence of the interaction the
density of dark energy will decrease up to a minimal value
corresponding to the state equation of dark energy
$\omega_{de}=-1$ and whereafter it will increase again as the
redshift becomes low. This can be thought as
 the phase transition of dark energy from the quintessence phase
to the phantom phase and the transformation point lies in $\omega
_{de}=-1$. The conclusion can be extended to the inverse that one
can deduce the the sign of $1+\omega_{de}$ from the evolutional
trend of the density of dark energy. A powerful probe of dark
energy is SNe Ia, which can be used as cosmological standard
cndles to measure how luminosity distance $d_{L}$ depends on
redshift in our universe. The relation between luminosity distance
and comoving distance $r(z)$ is:
\begin{equation}
d_{L}=(1+z)r(z)\,. \label{equ7}
\end{equation}
The density of dark energy can be extracted from the comoving
distance, as follows:
\begin{equation}
r(z)=H_{0}\int_{0}^{z}\frac{dz'}{H(z)}\,, \label{equ8}
\end{equation}
because the Hubble parameter depends on the density of dark energy
shown in Eq. (1). In the future more observational data will be
obtained. As long as we have some information about the change
rate of the density of dark energy we can deduce the sign of
$1+\omega_{de}$ which will provide some principles to choose the
appropriate model of dark energy. And if the phase transition is
confirmed in the future, the state equation of dark energy is
bound to evolve from $\omega_{de}>-1$ to $\omega_{de}<-1$.

In summary, in this paper we consider there are only the
nonrelativistic matter and dark energy in the universe. Ignoring
their interaction, the matter and dark energy satisfy their
conservation condition on their own. The change rate of the
density and the fractional density of dark energy are deduced from
the conservation condition. The fractional density of dark energy
increases as we go to the lower redshift and gradually dominates
the universe which has an important implication that it is not
significant to discuss the fractional density of dark energy
because as long as give a model of dark energy with negative
pressure in the absence of the interaction between the matter and
dark energy, it is certain to obtain the gradually increasing
fractional density of dark energy. We should focus on discussing
the density itself of dark energy rather than the fractional
density of dark energy.

Combining the prior of the state equation of dark energy being
below $-1$ around the present epoch evolved from $\omega _{de}>-1$
in the past, the density of dark energy will decrease to a minimum
before continually increases as the redshift becomes low which can
be thought as the phase transition of dark energy from the
quintessence phase to the phantom phase. On the base of these
conclusion there are many questions to settle, for example which
mechanism lead to the phase transition of dark energy, whether the
matter and dark energy exist the interaction or not and if in the
presence of the interaction how the density of dark energy will
change. We believe that we will make great strides in probing into
the nature of dark energy and the evolution of the cosmology by
settling these questions.

\begin{acknowledgments}
We are grateful to Mr. Lixin Xu for helpful discussions. This work
is supported by National Natural Science Foundation of China under
Grant NO.10573004.
\end{acknowledgments}

\end{document}